\documentclass[aps,prb,twocolumn,showpacs,amsmath,amssymb]{revtex4-1}

\usepackage{graphicx}
\usepackage{dcolumn}
\usepackage{bm}
\usepackage{color}
\usepackage{ulem}

\begin{document}

\title{
Multi-step atomic reaction enhanced by an atomic force microscope probe on Si(111) and Ge(111) surfaces
}

\author{Batnyam Enkhtaivan}
\author{Atsushi Oshiyama}

\affiliation{Department of Applied Physics, The University of Tokyo, Hongo, Tokyo 113-8656, Japan}

\begin{abstract}
We present first-principles total-energy electronic-structure calculations that provide the microscopic mechanism of the adatom interchange reaction on the Sn- and Pb-covered Ge(111)-(2$\times$8) and the Sb-covered Si(111)-(7$\times$7) surfaces with and without the tip of the atomic force microscope (AFM). We find that, without the presence of the AFM tip on the Ge surface, the adatom interchange occurs through the migration of the adatom, the spontaneous formation of the dimer structures of the two adatoms, the dimer-dimer structural transitions that induce the exchange of the positions of the two adatoms, and then the backward migration of the adatom. We also find that the dimer structure is unfeasible at room temperature on the Si surface and the adatom interchange are hereby unlikely. With the presence of the tip, we find that the reaction pathways are essentially the same for the Ge surface but that the energy barriers of the migration and the exchange processes are substantially reduced by the AFM tip. We further find that the AFM tip induces the spontaneous formation of the dimer structure even on the Si surface, hereby opening a channel of the interchange of the adatoms. Our calculations show that the bond formation between the AFM tip atom and the surface adatom is essential for the atom manipulation using the AFM tip.
\end{abstract}

\maketitle

\section{\label{sec:introduction}INTRODUCTION}

Atom-scale identification of structures of solid surfaces in real space has been achieved by the scanning probe microscopy in which either the tunneling electric current between the surface and the probing tip [scanning tunneling microscope (STM)\cite{stm}] or the forces acting on atoms on the tip [atomic force microscope (AFM)\cite{afm}] is measured at a certain lateral position with atomic resolution. This scanning probe technique has been also utilized to manipulate atoms on surfaces. Eigler and Schweizer first manipulated Xe atoms on Ni surfaces at 4 K,\cite{eigler} and then Crommie {\it et al.} made a quantum corral by placing Fe atoms on the Ni surface using an STM probe.\cite{crommie} The AFM probe has been also successfully utilized for the atomic manipulation on the Si surface,\cite{oyabu1} expanding the feasibility to non-conductive surfaces. Moreover, the manipulation of an atom on semiconductor surfaces at room temperature leads to the possibility of bottom-up fabrication processes in nanotechnology.

Since then, many types of atomic manipulation by the AFM probe has been achieved. They include vertical interchange of the tip and surface atoms on the Sn-covered Si(111) surface,\cite{vertical_interchange} lateral interchange of adatoms on the Ge(111)\cite{lateral_interchange1} and on the Si(111)\cite{lateral_interchange2} surfaces, lateral manipulation of single Si and Ge adatoms on the Si(111)\cite{lateral_shift1} and the Ge(111)\cite{lateral_shift2} surfaces, respectively, and the controlled diffusion of Ag adatom on the Si(111) surface.\cite{diffusion_gate} Albeit these stimulating experimental achievements, clarification of the underlying microscopic mechanisms of these atomic manipulation processes remains to be clarified.

The atomic manipulation is classified into two categories: lateral manipulation in which a surface atom is moved and placed at a certain site with the aid of the probe, and the vertical manipulation in which a surface atom is interchanged with a probe atom and then moved to a different position. As for the former, Sugimoto {\it et al.} first successfully achieved the lateral interchange of surface Sn and Ge adatoms on the Ge(111)-c(2$\times$8) surface\cite{lateral_interchange1} and then succeeded in lateral interchange of Sb and Si adatoms on the Si(111)-(7 $\times$ 7) surface.\cite{lateral_interchange2} It is noteworthy that such lateral interchange is observed without the probing tip: Ganz \textit{et al.} showed that Pb and Ge adatoms are interchanged with each other on the Ge(111)-c(2$\times$8) surface at temperatures from 24 to 79$^\circ$C without the probing tip.\cite{lateral_interchange3} This implies that the lateral interchange manipulation with the probing tip is essentially the enhancement of the combined process of the migration or diffusion and the subsequent exchange of adatoms. The aim of the present paper is to examine the possibility of these enhanced atomic reactions with the probing tip based on the first-principles calculations.

Atomistic calculations have provided insights into the interaction between the AFM tip and the surface atoms.\cite{afm_dft_perez1,afm_dft_perez2,afm_dft_dieska,jarvis_moriarty,dieska_stich1,dieska_stich2} Per\'{e}z {\it et al.}\cite{afm_dft_perez1,afm_dft_perez2} performed the total energy calculations based on the density functional theory (DFT)\cite{dft1,dft2} for Si tips on the Si(111) surface and have found that the covalent interaction between the tip and the surface atoms is important for the atom-scale resolution in AFM. It is also found that weak ionic interaction due to the charge transfer plays a role on the Cu(001) surface.\cite{afm_dft_dieska} The structure of the Si tip and its effect on the tip-surface interaction have been examined by the DFT calculation.\cite{jarvis_moriarty} As for the atom manipulation, Die\v{s}ka and \v{S}tich are first to provide a theoretical insight on the lateral interchange of Sn and Ge atoms on the Sn-covered Ge(111)-c(2$\times$8) surface.\cite{dieska_stich1,dieska_stich2} They have found that with the presence of an AFM tip, the energy barrier of the adatom migration is 0.6 - 0.8 eV but that the subsequent exchange is associated with the energy barrier of more than 1 eV, concluding that the lateral interchange is unlikely to occur. Then they proposed, based on the DFT calculation using specific tip structure model, that the interchange could occur via the exchange of the tip and the surface atoms. However in the corresponding experiment,\cite{lateral_interchange1} the discontinuity in the frequency shift which is usually expected in the atom transferring between the tip and the surface was not observed.

In this paper, we perform extensive density-functional calculations for the Sn-covered and Pb-covered Ge(111) surfaces and also Sb-covered Si(111) surface to resolve the microscopic mechanism for the lateral interchange of surface atoms. We first examine the diffusion (migration) and exchange processes without the AFM tip. We identify the diffusion pathways: There are several metastable atomic configurations and the diffusion takes place via step-by-step migration among the configurations. We find, when the target two atoms are close to each other, that the two atoms spontaneously form a dimer structure which provides a new reaction pathway for the exchange process. The energy barrier for the exchange process is substantially lower than that examined in the past.\cite{dieska_stich1,dieska_stich2}
We also find that the presence of the AFM tip lowers the migration barrier and further assists in the formation of the precursor dimer structure, leading to the more frequent lateral interchange.

The paper is organized as follows. The calculational methods and the pertinent conditions for the calculations are explained in Section \ref{sec:method}. 
The pathway of adatom migration and exchange processes on the Ge surface without the AFM tip is described in Sec.~\ref{subsec:notip_migration} and \ref{subsec:notip_dimer_exchange_ge}.
In Sec.~\ref{subsec:notip_dimer_exchange_si}, the adatom migration and exchange processes on the Si surface without the AFM tip is presented.
The modification of energy barrier of the adatom interchange process by the AFM tip is described in Sec.~\ref{subsec:tip_barrier_modification}.
Finally, we summarize our findings in Sec.~\ref{sec:summary}.

\section{\label{sec:method}CALCULATIONS}

Calculations are performed in the density functional theory (DFT)\cite{dft1,dft2} using Vienna Ab intio Simulation Package (VASP).\cite{vasp1,vasp2} The generalized gradient approximation (GGA)\cite{gga} is adopted for the exchange-correlation energy. Projector augmented-wave (PAW) potentials \cite{paw} are adopted to describe the electron-ion interaction. We use the cutoff energy of 200 eV for the plane-wave basis.

Each substrate surface is simulated by a repeating slab model. The atomic slab is separated from its adjacent image slabs by the vacuum region so that the atomic distances between the different slabs are more than 6 \text{\AA}, which is found to be large enough to neglect the interaction between the slab and its images. Each slab for the Ge(111)-c(2$\times$8) and Si(111)-(7$\times$7) surfaces is simulated by six atomic layers in addition to the adatom layer. The atoms at the bottommost layer of the slab are terminated with H atoms to remove unsuitable dangling bonds electronically. 
In the lateral directions,  Si(111)-(7$\times$7) surface is simulated by the dimer-adatom-stacking-fault (DAS) model of Takayanagi \textit{et al.}.\cite{das} A 6$\times$8 surface unit cell obtained by tripling the primitive cell of the Ge(111)-c(2$\times$8) surface in the short side direction is used to simulate the Ge(111)-c(2$\times$8) surface. Only $\Gamma$ point is sampled for the Brillouin zone integration for the supercell cells are large. Structural optimization is performed using calculated Hellmann-Feynman forces. All the atoms except for the bottommost layer atoms and the attached H atoms are relaxed until the forces acting on the atoms are smaller than 0.1 eV/\text{\AA}. 
The conditions explained above assure that the numerical error of the total energy is less than 0.04 eV.

To identify the reaction pathways of the interchange of surface atoms and the corresponding energy barriers, we adopt nudged elastic band (NEB) method.\cite{neb} This method partly assures continuity of the reaction pathway, compared with the hyperplane constraint method,\cite{jeong} by introducing fictitious elastic forces during the energy minimization.

%%%%%%%%%% FIGURE 1 %%%%%%%%%%%%%
\begin{figure}
\centering
\includegraphics[width=0.7\linewidth]{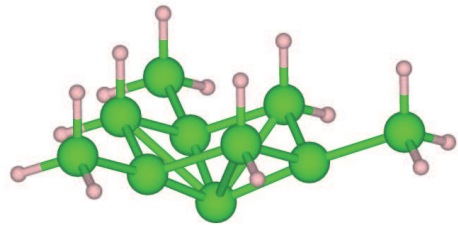}
\caption{
The model of the Si tip considered in the calculation. 
The green (large) and purple (small) spheres depict Si and H atoms, respectively. 
}
\label{fig:tip}
\end{figure}
%%%%%%%%%%%%%%%%%%%%%%%%%%%%

We consider the AFM tip composed of Si atoms and it is simulated by the atomistic model shown in Fig.~\ref{fig:tip}. The tip model consists of ten Si atoms and fifteen H atoms. This model is used in previous works.\cite{afm_dft_perez1,afm_dft_perez2,lateral_shift2,lateral_shift3,jarvis_moriarty,tip_apex_shape}
In our calculations, we have done structural minimization of this tip along with the surface atomic configurations. The H atoms and the Si atoms bonding with the H atoms in the tip, however, are fixed during the geometry optimization. It is noteworthy that the tip apex atom has a dangling bond which points toward the surface direction.

\section{\label{sec:results_and_discussion} RESULTS AND DISCUSSION}

In this section, we first present our results for the lateral interchange process without the AFM tip. We identify the detailed migration mechanism and find a pathway for the exchange process via a dimer structure newly found in our calculation. Then we show how the probe modifies the migration barriers and the exchange processes.

%%%%%%%%%% FIGURE 2 %%%%%%%%%%%
\begin{figure*}
\centering
\includegraphics[width=0.8\linewidth]{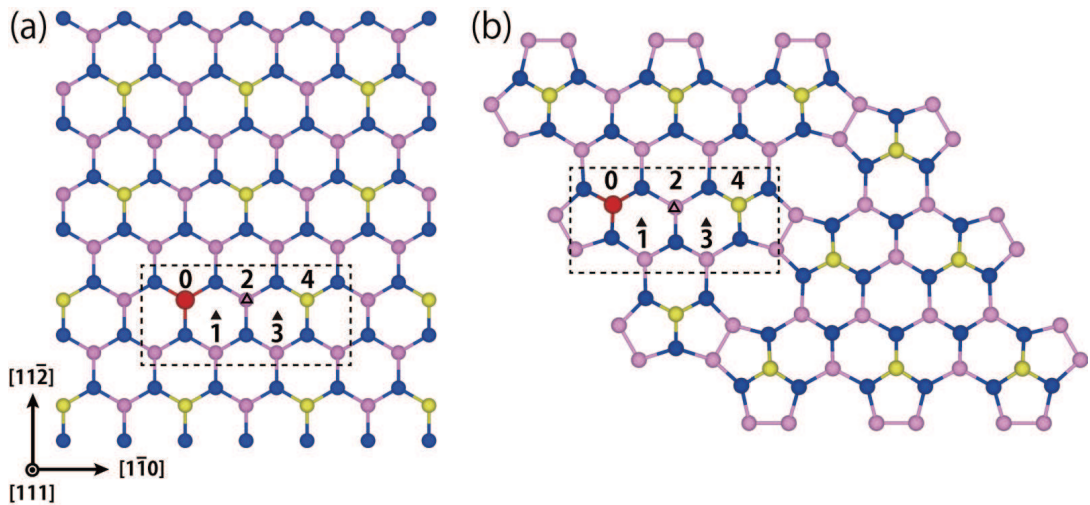}
\caption{
(Color online) Top views of (a) Ge(111)-c(2$\times$8) and (b) Si(111)-(7$\times$7) surfaces. In (a), the Ge adatoms and the second layer and third layer Ge atoms are shown by the yellow, blue and purple spheres, respectively. The foreign adatom such as Sn or Pb substituting for the Ge adatom is shown by the red ball. In (b), Si adatoms, the second-layer and the third-layer Si atoms are shown in yellow, blue and purple spheres, respectively. The adatom such as Sb substituting for the Si adatom is shown in red ball. The stable adsorption sites for the adatom, H$_3$ and T$_4$ (see text), are marked by $\blacktriangle$ and $\triangle$, respectively. Important sites during the interchange process are labeled by the numbers from 0 to 4 (see text). 
}
\label{fig:surfaces}
\end{figure*}
%%%%%%%%%%%%%%%%%%%%%%%%%%%

Figure \ref{fig:surfaces} shows schematic top views of Ge(111)-c(2$\times$8) and Si(111)-(7$\times$7) surfaces on which atomic interchange takes place. Experimentally, adatoms such as Sn and Pb are interchanged with a Ge adatom on the Ge(111)-c(2$\times$8) surface [Fig.~\ref{fig:surfaces}(a)] or Sb adatom is interchanged with a Si atom on Si(111)-(7$\times$7) [Fig.~\ref{fig:surfaces}(b)]. We examine the interchange process in which one of the two target atoms migrates near to the other atom for the exchange to take place, and then the other atom migrates to the vacant adatom site. There are two scenarios: In the first scenario, the foreign adatom at the site 0 in Fig.~\ref{fig:surfaces} migrates near to the host adatom located at the site 4; then when the two atoms are close enough to each other, an exchange process starts. After the exchange process, the adatom originally at site 4 migrates to the vacant site 0 ($I_1$ process hereafter). In the second scenario, the roles of the two atoms are changed: i.e., the host adatom at the site 4 migrates near the site 0 and the two atoms are exchanged, and then the foreign adatom originally at site 0 migrates to the vacant site 4 ($I_2$ process hereafter). It may be possible that both the two atoms migrate, meet together, are exchanged and then migrate back. Or the two atoms both migrate independently and reach the site where the partner atom is originally located. However, in these processes, the two sites at which the adatom disappears are generated. This costs the energy roughly twice compared with the situation in which one of the two atoms stay at the stable site. Hence we consider the two scenarios described above in this paper.

\subsection{\label{subsec:notip_migration} Adatom migration pathway and energy barrier on the Ge(111) surface without the tip }

%%%%%%%%%% FIGURE 3 %%%%%%%%%%%
\begin{figure*}[hbtp]
\centering
\includegraphics[width=0.7\linewidth]{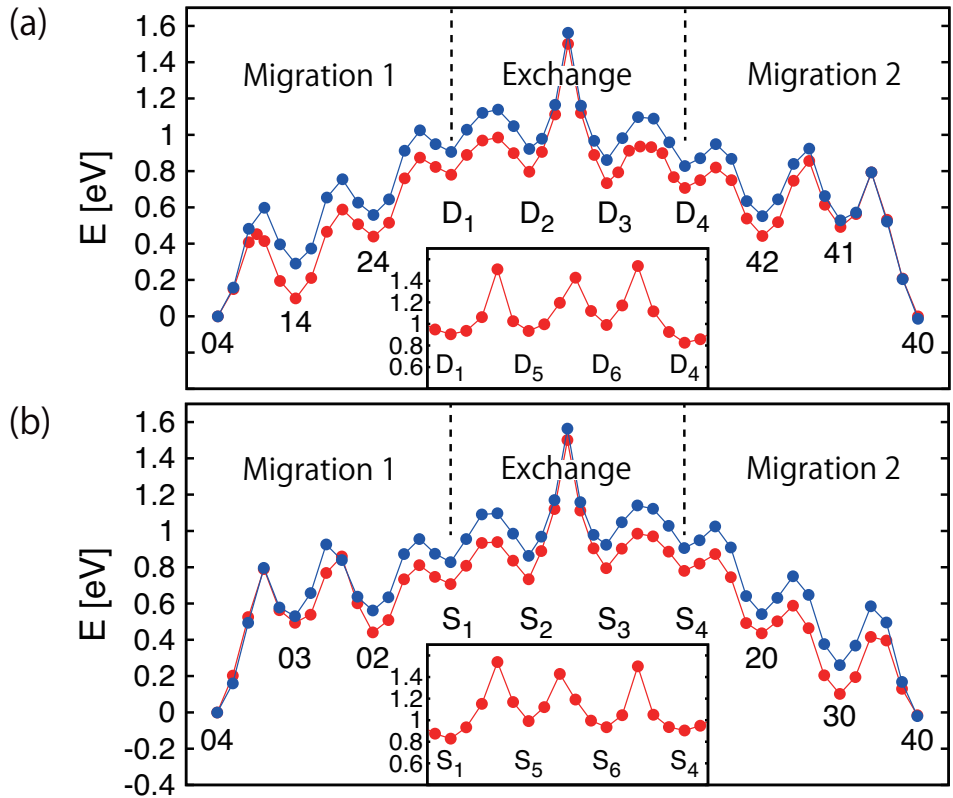}
\caption{
(Color online) 
Calculated energy profiles along the reaction coordinate for the atomic interchange between the Ge adatom and either the Sn (blue line) or the Pb (red line) foreign adatom on the Ge(111)-c(2$\times$8) surface. The interchange reaction is composed of the three steps: Either the Ge adatom or the foreign adatom migrates toward the other adatom (Migration 1), the exchange process via the dimer formation takes place (Exchange), and then the back migration of the other adatom takes place (Migration 2). In (a), the foreign adatom migrates toward the Ge adatom (the $I_1$ process), whereas in (b) the Ge adatom migrates toward the foreign adatom (the $I_2$ process). The metastable configurations during the reaction is labeled by two letters $nm$ in which the foreign adatom and the Ge adatom are located at sites $n$ and $m$ in Fig.~\ref{fig:surfaces}, respectively. The metastable configurations labeled by $D_k$ and $S_k$ depict the dimer configurations spontaneously formed (see Fig.~\ref{fig:dimer_strs} and text). In the insets, the energy profiles along the alternative pathways (path 2: see text) for the exchange of Sn and Ge are shown.  
}
\label{fig:i1i2}
\end{figure*}
%%%%%%%%%%%%%%%%%%%%%%%%%%%

In this subsection, we present the results of the migration pathway and the corresponding energy barrier in the interchange reaction on the Ge(111) surface without the tip. We find that the hollow sites labeled as H$_3$ and the on-top sites above the third-layer host atom labeled as T$_4$ (Fig.~\ref{fig:surfaces}) are (meta)stable for the adatom, in accordance with the previous work of Takeuchi {\it et al.}.\cite{takeuchi} At these sites, the adatom forms three bonds with the second layer Ge atoms which makes these sites (meta)stable.  

There are three steps in the adatom migration before the two adatoms get close to each other in $I_1$ and $I_2$ processes: In the $I_1$ process, the foreign adatom migrates from site 0 to site 1, then from site 1 to site 2, and thirdly from site 2 to site 3. Interestingly we have found that the foreign adatom spontaneously forms a dimer with the host adatom at site 3 (see below and Fig.~\ref{fig:dimer_strs}). We call this dimer structure $D_1$ hereafter. The energy profiles for the migration of Sn and Pb adatoms along the pathway we have determined are shown in the left part of the Fig.~\ref{fig:i1i2} (a), as labeled by Migration 1. During the three steps in Migration 1, the three (meta)stable configurations where the foreign adatom is located at site $n$ while the host adatom is located at site $m$. We call these configuration $nm$ hereafter (Fig.~\ref{fig:i1i2}). In Migration 1, the rate-determining processes are \textsl{04} $\rightarrow$ \textsl{14} and \textsl{14} $\rightarrow$ \textsl{24} migrations for Sn and Pb adatoms, respectively, as are deduced from our migration-energy calculations (Table \ref{tb:barriers}). The calculated rate-determining migration energy is 0.60 eV for Sn and 0.49 eV for Pb. This indicates that Pb migrates more frequently than Sn.

In the $I_2$ process, we have found that the host Ge adatom migrates toward the foreign adatom at site 0 via the three metastable configurations, \textsl{04}, \textsl{03} and \textsl{02}, as shown in Fig.~\ref{fig:i1i2} (b) (Migration 1). Again we have found that, after the \textsl{02} configuration, the Ge adatom forms the dimer structure spontaneously. In the Migration 1 process, the rate-determining process is the first movement from \textsl{04} to \textsl{03} (Table \ref{tb:barriers}). The calculated barriers are 0.80 eV for the Sn-covered surface and 0.79 eV for the Pb-covered surface. These migration barriers are definitely larger than those of the rate-determining barriers in the $I_1$ process, indicating that the foreign adatom is likely to move toward the host Ge adatom and then make the exchange.

After the two adatoms become close to each other, the exchange reaction via the dimer structures occurs (details are given in the next subsection). By the exchange, two adatoms change their places: e.g., $D_1$ and $D_4$ in Fig.~\ref{fig:dimer_strs}. After this exchange, one of the two adatoms migrates to the vacant site: Ge goes to the site 0 in the $I_1$ process and the foreign adatom goes to the site 4 in the $I_2$ process. The energy profiles of this migration ($D_4 \rightarrow \textsl{42} \rightarrow \textsl{41} \rightarrow \textsl{40}$ or $S_4 \rightarrow \textsl{20} \rightarrow \textsl{30} \rightarrow \textsl{40}$) are shown in Fig.~\ref{fig:i1i2}, as labeled by Migration 2.
We have found that the migration barriers in the Migration 2 process for Sn, Pb and Ge adatoms are relatively lower than those of the Migration 1 process (Table.~\ref{tb:barriers}). This is what is expected since the Migration 2 is the atomic transitions from the higher to the lower energy configurations whereas the Migration 1 is the contrary.

%%%%%%%%%%%% TABLE 1 %%%%%%%%%%%%%
\begin{table*}
\caption{\label{tb:barriers}
Calculated migration barriers and exchange barriers in the interchange reaction between the Ge adatom and either the Sn or the Pb foreign adatom on the Ge(111)-c(2$\times$8) surface. The values in the $I_1$ and $I_2$ processes are shown. The barriers are defined as the transition-state energy measured from the preceding metastable-state energy. The numbers 1 and 2 in the brackets in the first column depict the path 1 and path 2 (see text), respectively, in the exchange process. The subscripts, $k$ and $l$, denote one of the dimer structures during the exchange process (see Fig.~\ref{fig:i1i2} and text). 
}
\begin{ruledtabular}
\begin{tabular}{lccccccccc}
& \multicolumn{8}{c}{
Energy barrier (eV)}\\
\colrule
\textrm{$I_{1}$} &
 \textsl{04} $\rightarrow$ \textsl{14}  & \textsl{14} $\rightarrow$ \textsl{24} & \textsl{24} $\rightarrow$ $D_{1}$ & $D_{1}$ $\rightarrow$ $D_{k}$ & $D_{k}$ $\rightarrow$ $D_{l}$ & $D_{l}$ $\rightarrow$ $D_{4}$ & $D_{4}$ $\rightarrow$ \textsl{42} & \textsl{42} $\rightarrow$ \textsl{41} & \textsl{41} $\rightarrow$ \textsl{40}\\
\colrule
Sn (1)  &   0.60  &  0.47  &  0.47 & 0.23 & 0.64 & 0.24 & 0.12 & 0.37 & 0.26 \\
Sn (2)  &   -  &  -  &  - & 0.60 & 0.49 & 0.55 & - & - & - \\
Pb (1)  &   0.45  &  0.49  &  0.44 & 0.20 & 0.70 & 0.20 & 0.11 & 0.41 & 0.30 \\
\colrule
\textrm{$I_{2}$} &
 \textsl{04} $\rightarrow$ \textsl{03}  & \textsl{03} $\rightarrow$ \textsl{02} & \textsl{02} $\rightarrow$ $S_{1}$ & $S_{1}$ $\rightarrow$ $S_{k}$ & $S_{k}$ $\rightarrow$ $S_{l}$ & $S_{l}$ $\rightarrow$ $S_{4}$ & $S_{4}$ $\rightarrow$ \textsl{20} & \textsl{20} $\rightarrow$ \textsl{30} & \textsl{30} $\rightarrow$ \textsl{40}\\
\colrule
Sn (1)  &   0.80  &  0.40  &  0.39 & 0.26 & 0.70 & 0.22 & 0.12 & 0.21 & 0.32 \\
Sn (2)  &   -  &  -  &  - & 0.71 & 0.44 & 0.56 & - & - & - \\
Pb (1)  &   0.79  &  0.37  &  0.37 & 0.23 & 0.77 & 0.19 & 0.09 & 0.15 & 0.31 \\
\end{tabular}
\end{ruledtabular}
\end{table*}
%%%%%%%%%%%%%%%%%%%%%%%%%%%%%%%%%

\subsection{\label{subsec:notip_dimer_exchange_ge} Exchange process via dimer structure}

In this subsection, we explain how the two adatoms exchange their positions after the Migration 1 process. As stated above, we have found that the exchange becomes feasible via the formation of the dimer structures.

%%%%%%%%%% FIGURE 4 %%%%%%%%%%%
\begin{figure}[htbp]
\includegraphics[width=0.95\linewidth]{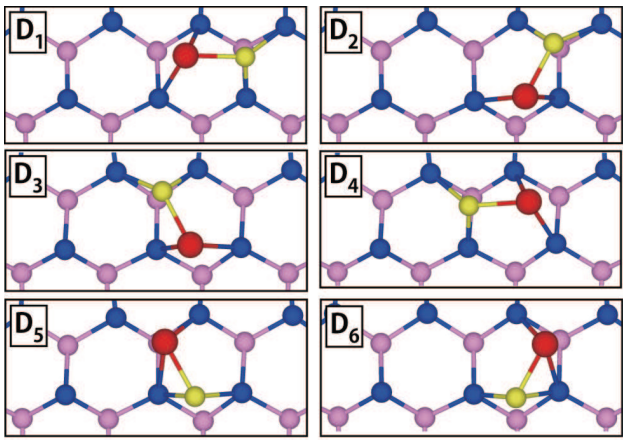}
\caption{
(Color online) 
Top views of the (meta)stable dimer structures during the exchange in the $I_1$ process on the Sn-covered Ge(111)-c(2$\times$8) surface. The red and the yellow balls depict Sn and Ge adatoms, respectively, which participate in the exchange. The region surrounded by dashed line in Fig.~\ref{fig:surfaces} (a) is enlarged. 
}
\label{fig:dimer_strs}
\end{figure}
%%%%%%%%%%%%%%%%%%%%%%%%%%%

After the Migration 1 in the $I_1$ process, the foreign adatom comes near to the site 3. However we have found that this site 3 is unstable against the formation of the dimer structure $D_1$ shown in Fig.~\ref{fig:dimer_strs}. We have also found other dimer structures, $D_2$, $D_3$, $D_4$, $D_5$ and $D_6$ shown in Fig.~\ref{fig:dimer_strs}, which are all stable and have nearly equal total energies. Then we expect the exchange process where several dimer structures appear sequentially. A possibility is a structural transition as $D_{1} \rightarrow D_{2} \rightarrow D_{3} \rightarrow D_{4}$ (path 1). An alternative way (path 2) is the transition as $D_{1} \rightarrow D_{5} \rightarrow D_{6} \rightarrow D_{4}$. 
We note that only a single adatom changes its position during this structural transition (Fig.~\ref{fig:dimer_strs}) so that the energy cost may be reduced. It can be seen that the positions of the two adatoms are exchanged during the $D_{1}$ and $D_{4}$ structures. For both Sn and Pb, this structural transition among the dimer structures takes place and the two adatoms are exchanged.

The calculated energy profiles for the exchange reactions in the $I_1$ process are shown in Fig.~\ref{fig:i1i2} (a) (the region labeled as Exchange). Corresponding energy barriers for each dimer-dimer structural transition are presented in Table~\ref{tb:barriers}. We have found that the energy barriers for the dimer-dimer transitions and thus the exchange reaction is substantially smaller than 1 eV for both paths 1 and 2: The rate determining process is $D_2 \rightarrow D_3$ with the barrier of 0.64 eV for the path 1 and $D_1 \rightarrow D_5$ with the barrier of 0.60 eV for the path 2. In both rate-determining dimer-dimer transitions, the Ge adatom rather than the foreign adatom hops to the nearby site. This is a consequence from the difference in the strength among the Sn-Ge, the Pb-Ge and the Ge-Ge bonds. These energy barriers are substantially lower than that for the direct exchange discussed in the past.\cite{dieska_stich2} The dimer formation we have found is the key process for the reduced energy barrier.

In the $I_2$ process, the Ge adatom comes near to the site 1 and then spontaneously forms a dimer with the foreign adatom ($S_1$ structure hereafter). We have also found other five (meta)stable dimer structures (from $S_2$ to $S_6$: not shown here) as in the $I_1$ process. Then the exchange reaction takes place through two paths: i.e., the $S_{1} \rightarrow S_{2} \rightarrow S_{3} \rightarrow S_{4}$ process (path 1) and the $S_{1} \rightarrow S_{5} \rightarrow S_{6} \rightarrow S_{4}$ process (path 2).
Sn or Pb adatom hops in $S_1 \rightarrow S_2$, $S_3 \rightarrow S_4$ and $S_5 \rightarrow S_6$, and Ge adatom hops in the rest of the dimer-dimer structural transitions. The calculated energy profiles for the exchange reactions in the $I_2$ process are shown in Fig.~\ref{fig:i1i2} (b) (the region labeled as Exchange), and the corresponding energy barriers for each dimer-dimer structural transition are presented in Table~\ref{tb:barriers}. We have found that the energy barriers for the dimer-dimer transitions and thus the exchange reaction are relatively low. Again the highest barriers of about 0.7 eV appear in the structural transitions in which the Ge adatom rather than the foreign adatom hops to the nearby site. The shape of the energy profile in the $I_2$ process is almost the mirror image of that of the $I_1$ process.

\subsection{\label{subsec:notip_dimer_exchange_si} Migration and exchange reactions without the tip on Si(111) surface} 

In this subsection, we examine the migration and exchange reactions of the Sb adatom on the Si(111) surface without the AFM tip. A schematic top view of Sb-covered Si(111)-(7$\times$7) is shown in Fig.~\ref{fig:surfaces} (b). By performing structure optimization calculations, we have found that $H_3$ and $T_4$ sites are (meta)stable adsorption sites as in the Ge(111) surface. Therefore, we consider the $I_1$ and $I_2$ processes in which either foreign adatom or the host Si adatom migrates through the metastable $H_3$ or $T_4$ sites. We label the sites by the numbers from 0 to 4 [Fig.~\ref{fig:surfaces} (b)].

%%%%%%%%%% FIGURE 5 %%%%%%%%%%%
\begin{figure}[htbp]
\centering
\includegraphics[width=0.8\linewidth]{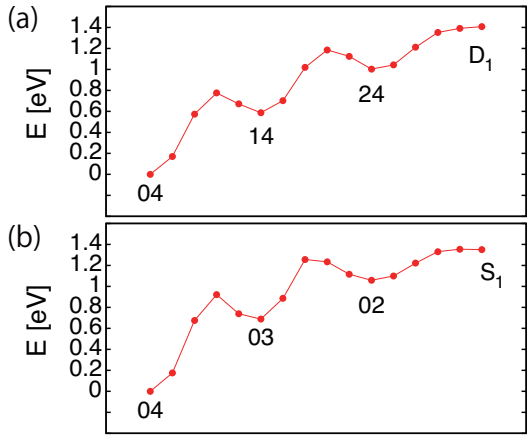}
\caption{
Calculated energy profiles along the reaction coordinate for (a) Sb adatom migration toward Si adatom and for (b) Si adatom migration toward Sb adatom on the Si(111)-(7$\times$7) surface. The $D_{1}$ and $S_{1}$ structures are the dimer structures having similar configurations to the $D_{1}$ structure shown in Fig.~\ref{fig:dimer_strs}.
}
\label{fig:sbsi}
\end{figure}
%%%%%%%%%%%%%%%%%%%%%%%%%%%%%%%%%%%%

The migration path of Sb adatom toward Si adatom is the same as that in Migration 1 in the $I_1$ process on the Ge(111) surface. The energy profile of this migration process, i.e., \textsl{04} $\rightarrow$ \textsl{14} $\rightarrow$ \textsl{24} $\rightarrow$ $D_{1}$, is shown in Fig.~\ref{fig:sbsi} (a). The rule of the labeling of the metastable configurations are the same as that in the previous subsections. The calculated migration barriers are 0.76 eV for the \textsl{04} $\rightarrow$ \textsl{14} process and 0.60 eV for the \textsl{14} $\rightarrow$ \textsl{24} process. After the migration, unlike the adatoms on the Ge(111) surface, Sb and Si adatoms do not form dimer structure: The $D_{1}$ structure shown in the energy profile is a structure similar to the $D_1$ structure of Ge and Sn adatoms on the Ge surface. However, it is not metastable structure and goes back to the structure \textsl{24} upon the structural optimization.

Similarly, in $I_2$ process, we have calculated the energy profile of Si adatom migration toward Sb adatom (the \textsl{04} $\rightarrow$ \textsl{03} $\rightarrow$ \textsl{02} $\rightarrow$ $S_{1}$ migration). The energy barriers for the \textsl{04} $\rightarrow$ \textsl{03} step and the \textsl{03} $\rightarrow$ \textsl{02} step are 0.92 eV and 0.57 eV, respectively, as shown in Fig.~\ref{fig:sbsi} (b). We have found that a dimer structure ($S_1$ structure) is metastable. However, it is not stable enough at room temperature for the energy barrier in the $S_1 \rightarrow \textsl{02}$ reaction is less than 10 meV.

The relatively high migration barriers of Sb and Si adatoms show that the migration of Sb and Si adatoms on the Si(111) surface is less frequent than that of Pb, Sn and Ge adatoms on Ge(111) surface. In addition to that, the dimer structure which assists in the exchange on the Ge(111) surface is practically unstable on the Si (111) surface, indicating that the adatom interchange is unlikely to occur on the Si(111) surface without the AFM tip.

\subsection{\label{subsec:tip_barrier_modification} Modification of diffusion barriers and exchange processes by the tip}

We are now in a position to examine the effect of the AFM tip on the interchange reaction between the foreign and host adatoms. We here present our result regarding the $I_1$ process on Ge(111) and Si(111) surfaces with the presence of the AFM tip. The time scale of the tip motion in the AFM measurements is in the order of microseconds, whereas that of the atomic motion on the surface is picoseconds. Hence we simulate the effect of the AFM tip in a static manner. The tip is placed at 5 \text{\AA} and 4.75 \text{\AA} distances from the surface for the Ge and Si surfaces, respectively. At these distances, the short-range attractive force between the tip and the surface is found to arise, which corresponds to the experimental situation. Here the tip-sample distance is defined as the vertical distance between the tip apex atom and the adatom on the unrelaxed surface.

In the experiments of adatom lateral interchange manipulation,\cite{lateral_interchange1,lateral_interchange2} AFM tip is scanned over the line connecting two adatoms with the time scale of milliseconds at fastest. To simulate this situation, tip apex atom is placed at the middle of the site 2 and the site 4 and then the interchange reaction is investigated.

%%%%%%%%%% FIGURE 6 %%%%%%%%%%%
\begin{figure*}[htbp]
\centering
\includegraphics[width=0.6\linewidth]{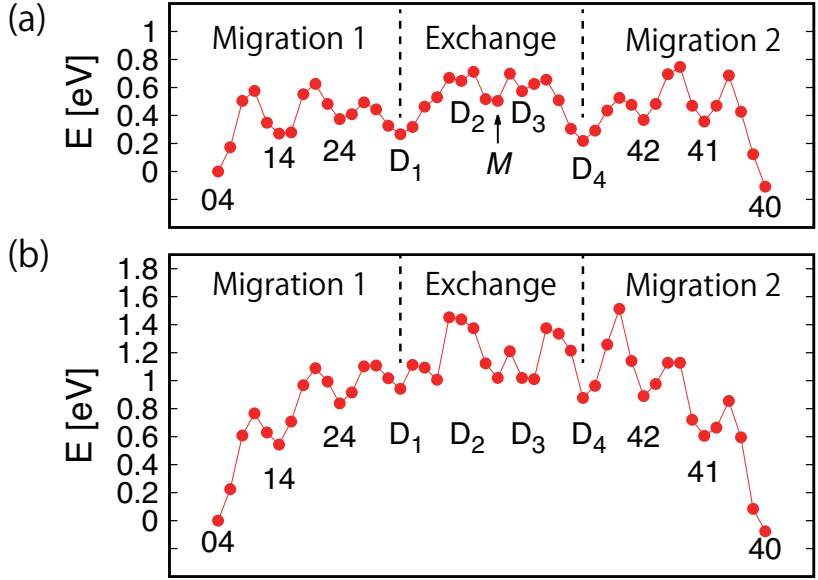}
\caption{
(Color online) 
(a) Calculated energy profile along the reaction coordinate for the atomic interchange (a) between the Ge adatom and the Sn adatom on the Ge(111)-c(2$\times$8) surface, and (b) between the Sb adatom and the Si adatom on the Si(111)-(7$\times$7) surface with the presence of the Si AFM tip. The interchange reaction is composed of the three steps: The Sn (Sb) adatom migrates toward the Ge (Si) adatom (Migration 1), the exchange process via the dimer formation takes place (Exchange), and then the migration of the Ge (Si) adatom (Migration 2). The labelling of the metastable configurations and the dimer structures is the same as in Fig.\ref{fig:i1i2}. 
}
\label{fig:tip_profile}
\end{figure*}
%%%%%%%%%%%%%%%%%%%%%%%%%%%

\subsubsection{Migration and exchange reactions on the Ge(111) surface with the AFM tip}

We have identified the migration and the exchange reaction pathways in the $I_1$ process on the Sn-covered Ge(111) surface. We have found that the reaction pathway is essentially identical to that with the absence of the AFM tip. However, the AFM tip substantially modulates the energy profile along the reaction coordinate. The calculated energy profile of the reaction with the presence of the AFM tip is shown in Fig.~\ref{fig:tip_profile} (a). The migration barrier from the structure \textsl{04} to the structure \textsl{14} is unchanged by the presence of the AFM tip (cf. Fig.~\ref{fig:i1i2} (a)). However, the barrier of the \textsl{14} $\rightarrow$ \textsl{24} migration is lowered from 0.47 eV to 0.35 eV by the AFM tip. The lateral distances between tip apex atom and Sn atom in the saddle point configuration is about 3.0 \text{\AA}. It shows that the interaction between the tip apex atom and the migrating adatom starts from relatively large distance. Not only the migration barrier is lowered, but also the total energy difference between the most stable \textsl{04} and the metastable \textsl{24} configurations is decreased from 0.56 eV to 0.37 eV. This means that the \textsl{24} configuration is stabilized by the AFM tip.

The exchange process is also enhanced by the AFM tip. As stated above, the rate-determining exchange process without the tip is the $D_2 \rightarrow D_3$ with the barrier of 0.64 eV for the path 1 and the $D_1 \rightarrow D_5$ with the barrier of 0.60 eV. With the presence of the AFM tip, we have found that the transition state between the $D_2$ and the $D_3$ in path 1 for instance disappears with the presence of the AFM tip. Instead, an energy dip labeled by \textit{M} emerges [Fig.~\ref{fig:tip_profile}(a)]. This is a consequence of the formation of the covalent bond between the tip apex atom and the hopping Ge atom (see below). As a result, the rate-determining process becomes the $D_1 \rightarrow D_2$ dimer-dimer transition with the barrier of 0.40 eV [Fig.~\ref{fig:tip_profile} (a)].

In the path 2, the tip apex atom forms a bond with the Sn adatom and pins it at the position in the $D_1$ structure. Consequently, the hopping of Ge adatom is suppressed in $D_1 \rightarrow D_5$ transition with the barrier of 0.89 eV due to the restricted relaxation of the Sn adatom. This implies that the enhancement and the suppression of the reactions generally depend on the atom-scale positioning of the AFM tip.

The energy profile for the interchange reaction obtained by our GGA calculations unequivocally clarifies that the AFM tip reduces the energy barrier in both migration and the exchange processes.

\begin{figure*}[htbp]
\centering
\includegraphics[width=0.7\linewidth]{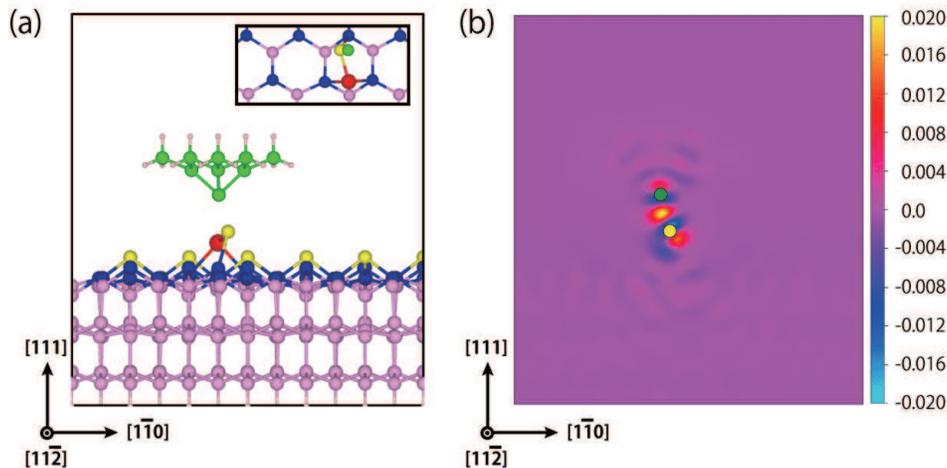}
\caption{
(Color online) 
(a) The atomic configuration of the \textit{M} (Fig.~\ref{fig:tip_profile} (a)). In the inset, the top view of a part of the surface region is shown. The size and the coloring of the balls are the same as those of Fig.~\ref{fig:tip} and Fig.~\ref{fig:surfaces}. (b) The charge density difference $\Delta \rho(\bm{\mathrm{r}})$ on the (11$\bar{2}$) plane. The plotted data is obtained by integrating $\Delta \rho(\bm{\mathrm{r}})$ along the [11$\bar{2}$] direction at each point of the (11$\bar{2}$) plane. The green and the yellow circles show the position of the tip apex atom and the Ge adatom, respectively.
}
\label{fig:charge}
\end{figure*}

In order to clarify the microscopic reason for the reduction of the reaction barriers, we have examined the charge density at the \textit{M} configuration during the $D_2 \rightarrow D_3$ structural transition. Fig.~\ref{fig:charge} shows the structure and the calculated charge density difference $\Delta \rho(\bm{\mathrm{r}}) = \rho(\bm{\mathrm{r}})-[\rho_{\textrm{s}}(\bm{\mathrm{r}}) + \rho_{\textrm{t}}(\bm{\mathrm{r}})]$, where $\rho(\bm{\mathrm{r}})$, $\rho_{\textrm{s}} (\bm{\mathrm{r}})$, and $\rho_{\textrm{t}} (\bm{\mathrm{r}})$ are the charge densities of the tip-surface system at the \textit{M} configuration, the isolated tip, and the surface without the tip, respectively. The tip apex Si atom located above the Ge adatom forms a bond with the Ge, which is evidenced by the bond charge shown in Fig.~\ref{fig:charge}(b). This bond formation makes the \textit{M} configuration energetically favorable and thus assists in the hopping of the Ge adatom during the $D_2 \rightarrow D_3$ structural transition.

%%%%%%%%%% FIGURE 7 %%%%%%%%%%%
%%%%%%%%%%%%%%%%%%%%%%%%%%%

\subsubsection{Migration and exchange reactions on the Si surface with the AFM tip}

The tip-surface interaction not only reduces the energy barriers in the reactions but also opens an otherwise unfeasible reaction pathway. As stated above, the dimer structures of the two adatoms are impossible on the Si(111) surface without the AFM tip, thus making the exchange reaction unfeasible. In contrast to that, we have found that the dimer structure which is a precursors for the exchange reaction is formed with the presence of the AFM tip on the Si(111) surface. Further we have found the four dimer structures, $D_1$, $D_2$, $D_3$ and $D_4$ as on the Ge surface, on the Sb covered Si(111)-(7$\times$7) surface. Then the pathway for the interchange reaction opens also on the Si(111) surface with the AFM tip. The energy profile of $I_{1}$ process in the interchange reaction of Sb and Si adatoms on Si(111)-(7$\times$7) surface with the presence of the AFM tip is shown in Fig.~\ref{fig:tip_profile} (b). The energy for the $\textsl{04} \rightarrow \textsl{14}$ and $\textsl{14} \rightarrow \textsl{24}$ migrations are 0.77 eV and 0.55 eV, respectively. Similar to the adatom migration process on the Ge surface, the barrier of $\textsl{14} \rightarrow \textsl{24}$ migration is lowered from 0.60 eV to 0.55 eV and the $D_{1}$ dimer structure is stabilized due to the interaction of the adatom with the AFM tip. The energy barriers among the dimer-dimer structural transitions are in the range of 0.2 eV - 0.5 eV. In the whole interchange reaction, the rate-determining processes are the migration \textsl{04} $\rightarrow$ \textsl{14} and the dimer dissociation $D_{4}$ $\rightarrow$ \textsl{42} with the energy barriers of 0.68 eV and 0.63 eV, respectively. The total energy difference of the structures \textsl{04} and \textsl{40} as observed in Fig.~\ref{fig:tip_profile} (a) comes from the difference in the interaction of Si and Sb adatoms with the tip apex atom. 

\section{\label{sec:summary}Summary}

We have performed total-energy electronic-structure calculations using density functional theory for the adatom interchange reaction on the Ge(111) and Si(111) surfaces with and without the probing tip of the atomic force microscope (AFM). We have first clarified the atom-scale reaction pathways for the adatom interchange without the AFM tip. We have found that the interchange reaction on the Ge(111) surface occurs as follows: (i) an adatom migrates toward the other adatom (forward migration), (ii) it spontaneously forms a dimer with the other adatom, (iii) there are several total-energy equal dimer structures, and the exchange of the two adatoms is realized through the dimer-dimer structural transitions, and finally (iv) the exchanged adatom migrates to the vacant site generated by the forward migration (backward migration). The calculated energy barriers are substantially lower than the barriers discussed in the past. This is due to the multistability of the dimer structures newly found in the present calculation. On the Si(111) surface, on the other hand, we have found that the dime structures are not spontaneously formed and the dimer-assisted interchange reaction is hereby impossible, leaving the interchange unfeasible. We have also identified the reaction pathways and the corresponding energy barriers for the interchange reaction with the presence of the AFM tip. For the Ge(111) surface, we have found that the reaction takes place similarly, i.e., the forward migration, the dimer-assisted exchange and the backward migration. Further, due to the bond formation of the tip apex atom and the surface adatom, the reaction barriers for the migration and the exchange processes are substantially reduced. We have also found that the AFM tip induces qualitative difference in the interchange phenomena on the Si(111) surface. The reaction channel for the exchange through the dimer formation opens with the tip-surface interaction: We have found that the formation of the covalent bond between the tip apex atom and the surface adatom causes the formation of the multistable dimer structures and thereby make the interchange reaction feasible. Our calculations show that the bond formation between the AFM tip atom and the surface adatom is the essential physics for the atom manipulation. This also implies that the precise placement of the AFM tip is important in more sophisticated atomic manipulation.

\begin{acknowledgments}
This work was supported by the ``Computational Materials Science Initiative'' and also by the project for Priority Issue (Creation of new functional devices and high-performance materials to support next-generation industries) to be tackled by using Post `K' Computer, both conducted by Ministry of Education, Culture, Sports, Science and Technology, Japan. Computations were performed mainly at the Supercomputer Center at the Institute for Solid State Physics, The University of Tokyo, The Research Center for Computational Science, National Institutes of Natural Sciences, and the Center for Computational Science, University of Tsukuba.
\end{acknowledgments}

\end{document}